# A prospectus on the surface metrology of seborrheic keratoses

**Authors**
Nicole Werpachowski[1] : nicole.werpachowski@gmail.com
Juliette Nutovits[1] : jnutovit@nyit.edu
Therese Limbana[1] : therese.anne15@gmail.com
John Goncalves[1] : jgonca03@nyit.edu
Alina Bridges[2] : alinabridges@icloud.com
Brian Lee Beatty[1] : bbeatty@nyit.edu

[1] New York Institute of Technology College of Osteopathic Medicine, Northern Blvd, Old Westbury, New York 11568, USA
[2] Dermatology Department, Donald and Barbara Zucker School of Medicine at Hofstra Northwell, New Hyde Park, NY, USA

**Corresponding Author**: Brian L. Beatty, New York Institute of Technology College of Osteopathic Medicine, Department of Anatomy, Old Westbury, New York 11568; email: phone: 516 686 7435; fax: 516 686 3740; email: bbeatty@nyit.edu

ORCID ID: Beatty 0000-0002-5464-0041

# The surface metrology of seborrheic keratoses

Nicole Werpachowski[1], Juliette Nutovits[1], Therese Limbana[1], John Goncalves[1], Alina Bridges[2], Brian Lee Beatty[1]

N.W., J.N., T.L., should be considered joint first authors.

[1]NYIT College of Osteopathic Medicine, Northern Blvd, Old Westbury, New York 11568, USA
[2]Dermatology Department, Donald and Barbara Zucker School of Medicine at Hofstra Northwell, New Hyde Park, NY, USA

Author Contribution:
N.W., J.N., T.L. should be considered joint first authors. N.W., J.N., T.L., J.G., B.B. designed the research study. N.W., J.N., T.L., B.B., performed the research. B.B. contributed the essential equipment. A.B performed the histopathology. N.W., J.N., T.L., B.B. analyzed the data. N.W., J.N., T.L., B.B., A.B wrote the paper. N.W., J.N., T.L., J.G., B.B., A.B read and approved the final manuscript.

Acknowledgements
Kelsi Hurdle provided support and access to the Sensofar S Neox at the NYIT College of Osteopathic Medicine Visualization Center.

Conflict of Interest:
The authors declare no conflicts of interest.

Abstract
**Background**: Human skin texture has yet to be quantified for diagnostic purposes. Here, the surface metrology of seborrheic keratoses is investigated with an optical profiler. **Materials and Methods**: Dermatologic specimens of 7 cadavers were prepared. Specimens were molded with polyvinyl siloxane and casts prepared with resin, which were scanned using a 3D white light optical profiler. Each scan produced 48 variables, categorized into 3 groups for each location: control, lesion center, and lesion edge. Images of the histopathology slides for suspected seborrheic keratoses were reviewed by a dermatopathologist. **Results:** The parameters under investigation included border versus center of keratoses, age, sex, lesion location, degree of sun exposure, and cause of death. Although some parameters differ between individuals and age groups, the majority of differences identified between the roughness parameters measured are a result of sex, sun exposure, and histological diagnosis, listed in order of increasing importance. Histological diagnosis provided the most significant, definitive number of individual measurements of areas of roughness in seborrheic keratoses in comparison to other parts of the skin (regardless if those were controls or a different pigmented lesion aside from seborrheic keratoses). **Conclusions:** This study demonstrates that there are quantifiable patterns of surface textures that can be compared between keratotic and standard normal skin surfaces. These findings suggest this method has the potential to be applied as a noninvasive adjunct to current methods of dermatological diagnosis.

1 | INTRODUCTION

The advancement of early diagnosis of dermatopathologies with noninvasive technologies such as high-frequency ultrasound, dermoscopy, and reflectance confocal microscopy has been extensively demonstrated.[1-2] Although these techniques have proven invaluable in their use of diagnosing without biopsying,[3-9] their visual nature captures data more akin to a histological section. This is a challenging task for experts to make diagnoses with, as many of the lesions appear similar on planar view.

Rather than exclusively using visual data for diagnostic purposes, texture can be used as a complementing modality.[10-12] Texture remains valuable in diagnosing, measuring degree of skin involved with tumor growth, chronicity of growth, and disorder of growth.[13] However, only a few of the current technologies measure topographical detail.[1] This leaves topographical diagnostics primarily in the hands of palpation, and tactile sensation plays a crucial role in the clinical examination of skin, especially in the field of dermatology. Dermatologists rely on visual cues and palpation to assess tactile descriptive information, including size and texture of tumor growth, but clinical accuracy is not absolute, highlighting the potential for diagnostic error.[13,41] Furthermore, palpation as a diagnostic tool may not be feasible in circumstances where there is risk of secondary infection or potential for causing damage or alteration to the affected area.[14] Therefore, while texture analysis offers significant advantages in assessing skin dermatoses, its current limitations underscore the need for ongoing development of technologies that can accurately capture topography of skin.

Surface texture has successfully become quantifiable for soft items with the advent of non-contact surface metrology methods. In comparison to *in vivo* reflectance confocal microscopy studies,[1-9] surface metrology is not merely a scanning method leading to visual images. Rather, it is used to produce a point cloud defining a surface and then quantify that surface in extreme mathematical detail.[38,39] These data can then be used to search for patterns of geometry that might elude visual observation or palpation, and can be done at resolutions resembling visual or palpatory scales as well as scales well below the limits of those normal human sensory modalities. Though initially developed primarily to measure worn surfaces of hard materials in machines, surface metrology has increasingly become used for the quantitative study of hard and soft biological tissues.[15-17] This strengthens our rationale to employ surface metrology in our study. As palpatory skills are not uniform for physicians,[18] nor is the spatial density of

Meissner's corpuscles,[19] surface metrology can not only be useful as a quantitative version of palpatory techniques, but also as a means of achieving uniformity in regard to the assessment of surface texture.

Seborrheic keratoses (SKs) are one of the most common epidermal tumors and are the most common benign skin tumor in the elderly population. They are a proliferation of immature keratinocytes believed to arise from a genetic or a sporadic component. They are distinguished by several unique surface features, including milia-like cysts, comedo-like openings, and a greasy, lobulated surface. Some SKs can be pedunculated with crusts and thus may have features of malignant lesions, making the diagnosis more complicated.[34,35] SKs are most commonly found on the trunk, back of hands, forearms, head, face, and neck and are known to increase in size and number with age.[27] Despite their benign nature and reassuring lack of transformative potential to skin malignancy, SKs are often deemed not therapeutically important to many clinicians.[29] However, patients who present with multiple SKs face a higher risk of having a lesion mistakenly identified as an SK due to features that could clinically mimic other cutaneous lesions or neoplasms, including basal cell carcinoma[30] and melanoma[31], due to their pigmentation and papillomatous, greasy appearance.[33]

Different approaches are used to gather and analyze data about SKs, all of which help in the identification and management. To aid in their diagnosis, SKs have been thoroughly measured using tools including dermatoscopy, ultrasound, and in vivo reflectance confocal microscopy.[4,20-26] Dermoscopy is a basic approach through which the clinician is able to assess surface features.[33] Dermoscopic analysis has been pivotal in identifying key features in SK-like melanomas, highlighting the value of detailed surface assessments.[33] It is the preferred non-diagnostic method for the identification of SKs in practice due to its accessibility and efficiency in dermatology. High resolution ultrasound, another noninvasive diagnostic modality, has captured appreciable differences in the acoustic characteristics of SKs and other pigmented skin tumors, noting that SKs have increased attenuation compared to melanomas, which helps in differentiation.[34]

However, there are some limitations to these established tools. For instance, a serious limitation of visual examination with dermatoscope is a large number of false positives, leading to unnecessary biopsies or removal, higher cost of treatment, and negative patient outcomes.[40] Another study noted that a significant drawback of diagnosing SKs with reflectance confocal microscopy (RCM) is related to variability in both clinical and RCM appearances of SKs and limited depth of RCM.[28] Thus, the capability of diagnostic techniques being able to differentiate SKs from other malignant lesions is essential, illustrating the need for more sophisticated imaging in dermatology. Furthermore, human skin texture has yet to be quantified for diagnostic purposes. Our study aims to quantify and delineate surface parameters of SKs in comparison to normal skin as an introduction to a further topical measurement regarding dermatopathies and their diagnosis.

What can be expected of surface metrology of skin in this way? SKs should have raised surfaces that make for greater relief between the edges and the center, along with edges with more extreme relief in comparison to nearby healthy skin. These differences in relief should manifest amid the height variables most, resulting in greater absolute values. SKs should also include much more irregularity in areal roughness, with much less uniformity of orientations of faces and greater complexity of surfaces. SKs are prone to becoming waxy or scaly on morphology, and this should result in major differences in surface complexity and numbers of individual peaks and valleys. It is the goal of this research to quantify and more precisely delineate these important texture parameters that aid in SK identification.[13] The parameters examined with the optical white light profiler could provide a compelling case for utilizing this method to study other benign and/or malignant skin pathologies by comparing their different surfaces and quantifiable textures.

## 2 | MATERIALS AND METHODS

Dermatological specimens of suspected seborrheic keratoses (SKs) and control sample of skin from the lateral thigh were prepared from seven whole body donor cadavers from the collection of the New York Institute College of Osteopathic Medicine Anatomy Lab used for teaching purposes. This collection is licensed by the New York State Department of Health with a Non-transplant Anatomic Parts license, including dermatological specimens. The identities of donors are recorded, but their identities aside from age, gender, sun exposure, and cause of death have been protected for this study. Each identified dermatological specimen was carefully washed with soap and water and given time to air dry, allowing preservation of the natural skin surface. The following procedure mirrors a study that investigated the surface metrology of bone surface attachments of knee ligaments.[15]

### 2.1 | Imaging, molding, casting, and histopathology procedure

Specimens of SKs and normal control skin were photographed with a Dino-Lite Edge digital microscope (Dino-Lite US | Dunwell Tech, Inc.) (Figure 1) and then molded using Polyvinylsiloxane molding material (President Jet normal body, manufactured by Coltene Inc.), applying it to the entire surface of the lesion and/or control skin. This molding material is the standard for dental microwear studies due to its features of setting quickly, separating clean from molded surfaces, and retaining a high resolution.[32] Molds hardened for ten minutes and were then removed; care was taken to avoid contamination of surfaces during removal and storage. Afterwards, each mold had a wall of polyvinylsiloxane putty built around it to improve resin retention and prevent spillage for casting. Casts were made with Epokwik clear epoxy resin

(Buehler), which was degassed using a vacuum chamber and poured into the molds, and then exposed to hand crank centrifuge to remove any additional bubbles. Casts then cured for 24 hours.[15]

Each suspected SK lesion specimen was also prepared for histopathology analysis. Once molds were removed, each lesion was dissected free from the surrounding skin with a scalpel, carefully done to avoid damage to the region of interest. Then each was cut into slices through the epidermis and dermis, and into the hypodermis, such that a section perpendicular to the normal plane of the skin was excised at 2-3 mm in thickness and placed into a standard histology tissue cassette. Samples of this nature were taken from the center of each lesion, and if lesions were large enough, a second section was taken from the edge of the lesion. Cassettes were then processed in an automated paraffin embedding machine (Sakura Tissue-Tek VIP Tissue Processor), then positioned in paraffin wax blocks (Arcadia Embedding Station) for sectioning on a microtome (Leica). Slides were then stained with hematoxylin and eosin and scanned with a slide scanner (Zeiss AxioScan 7). Images were saved as .czi files and converted to TIFF and PNG for easier data management.

2.2 | Surface scanning procedure (orientation, parameters, etc.)

Each cast was scanned with a Sensofar S Neox white light optical profiler at the NYIT College of Osteopathic Medicine Visualization Center (NYIT, Old Westbury, NY). The scanning procedure was identical for each specimen. Casts were oriented on the scanner so that the approximate center or border of the lesion specimen was at the center of the viewing/scanning rectangle. Each cast was scanned at x5 and x20 magnification separately, leaving scanned areas of 3.49mm x 2.63m and 873.33$\mu$m x 656.61$\mu$m, respectively. The z range was defined by what was in focus and the scan was executed. Scans whose data completeness was equal to or greater than 99% were included in the study (Figure 2).

2.3 | Postprocessing scan data

Files of the scans were imported into SensoMAP software and subject to a series of processing steps including leveling, retouching, refilling non-measured points, removing form (polynomial = 3), and repeat leveling. Retouching is intended to edit out contaminants and unusual reflections/spikes, including those resulting from bubbles or dust particles. Refilling non-measured points replaces gaps (areas not scanned or deleted during retouching), creating a surface continuous with surrounding points. Removing form minimizes the unintended consequences of scanning and measuring warped surfaces.[15] After completing this series of processing steps, data created and extracted from each scan included all available ISO-25178-2 parameters, texture direction, motifs analysis, and scale-sensitive fractal analysis (Figure 3). All data was saved to an Excel spreadsheet.

Images of the histopathology slides for suspected SK lesions were reviewed by a board-certified dermatologist and dermatopathologist (Figure 4). These determinations were used to sort the suspected lesions into SK vs non-SK. Only the histopathologically confirmed SK lesions were used as data and analyzed statistically using IBM SPSS Statistics (version 28).

2.4 | Grouping and variables

48 variables were examined, each of which quantified surface texture uniquely, in the context of distinct groups: ISO 25718-2 height parameters, hybrid parameters, functional parameters, functional (volume) parameters, feature parameters, and functional parameters (stratified surfaces), as well as texture direction measures, motifs analysis, and scale-sensitive fractal analysis (Table 1). For more on these parameters, see other recent studies that have utilized them.[15,39,59] All parameters available with the optical white light profiler programming were used. The parameters under investigation in this study were SK lesion border versus center, age, sex (male or female), location of lesion, level of sun exposure (categorized by none, low level, or high level) and cause of death.

3 | RESULTS

3.1 | ANOVA and t-tests

First, a few tests were done to see if any surface metrology parameters differed for controlled or spurious reasons. The most obvious is to look for differences between control regions and the border and center regions scanned. When subject to ANOVA, the only differences found exist between control regions and the border (*Ssk* $p = 0.048$; *number of motifs* $p = 0.009$) or control versus center regions (*Ssk* $p = 0.048$; *Spd* $p = 0.011$; *number of motifs* $p = 0.009$). It makes sense that border regions would muddle these comparisons, as they contain the lesion and adjacent, healthy skin. The *number of motifs* is sensitive to perturbations of measurable separate regions of elevation, which would naturally differ between all three regions.

With a large number of samples from a limited number of individuals, it is important to identify if individual differences might account for significant differences, whether because of the nature of their preservation or living individual variation. If this variation had been more significant, one might reconsider using these measures to investigate other associated patterns. The fact that only *Sku* ($p = 0.008$) was significantly different relieves us of that concern, especially as Tukey's post hoc test indicates that this is entirely because of differences between individual 2 with 3, 4, 5, 6, and 7. Each of the seven individuals were different ages, and it is worth noting that *Sku* ($p = 0.006$) is the parameter that differs due to age, with Tukey's post hoc test identifying the source of that significance as lying between individual 1 (age 70) and individuals 2 (age 81), 3 (age 94), 5 (age 97) and 6 102), but NOT different from individual 4 (age 95). These results about individuals and ages suggest that *Sku* differences are the result of spurious differences among individuals.

Among these individual differences, sex might have a more expected significance in differences, largely because of tendencies toward better skin care in females and known existing skin textural differences between the sexes. The t-test for significance between the sexes found a large number of parameters significantly different between males and females: *Sq* (p = 0.032), *Sp* (p = 0.021), *Sv* (p = 0.007), *Sz* (p = 0.009), *Sa* (p = 0.041), *Smc* (p = 0.049), *Sxp* (p = 0.034), *S10z* (p = 0.012), *S5p* (p = 0.018), *S5v* (p = 0.012), *Sk* (p = 0.019), *Spk* (p = 0.016), *Svk* (p = 0.015), and *Smfc* (p = 0.010). Studies have not found significant gender differences in relation to the incidence of SKs aside from reporting observations of different frequency of SK in different locations between the sexes.[24] While some studies have demonstrated skin texture can be used to differentiate males from females[47], other studies of skin texture do not support these conclusions[48]. The purpose of this study was not to use skin texture to discriminate sexes, but this is a worthwhile avenue to pursue further that our data point to.

Another nuanced difference between these individuals comes from the limited data we have on their cause of death (according to their death certificates). Like with individual identity and age, *Sku* (p = 0.004) is the only significantly different variable, and Tukey's post hoc test points to these differences being due to differences between individuals with cancer versus heart disease, cancer versus infarction, and cancer versus unknown cause of death. No differences were found between any other pairings, or anyone with respiratory failure with these other groups. In contrast, when cause of death was simplified to cardiovascular disease or the broader category of "other", a few more parameters were significantly different: *Sku* (p < 0.01), *Smr* (p = 0.016), and *Height mean* (p = 0.013).

ANOVA results attempting to distinguish patterns of skin texture between body regions exhibit no significant differences between regions, whether testing specifically the chest/back/upper extremity/lower extremity or the simplified approach of torso/limb. This might not be the distinguishing characteristic that matters most, as the t-test discerning sun exposure levels appears to have found many parameters with significant differences: *Sp* (p = 0.025), *Sv* (p = 0.049), *Sz* (p = 0.032), *Sal* (p = 0.020), *Vm* (p = 0.023), *Vv* (p = 0.025), *Vmp* (p = 0.023), *Vmc* (p = 0.023), *Vvc* (p = 0.025), *Vvv* (p = 0.031), *Spc* (p < 0.001), *S10z* (p = 0.022), *S5p* (p = 0.006), *Sda* (p = 0.017), *Sha* (p = 0.017), *Sdv* (p = 0.017), *Shv* (p = 0.017), *Sk* (p = 0.036), *Area mean* (p = 0.017).

Lastly, when categorized by histopathological diagnosis done by a board-certified dermatologist (author AB), a much greater number of parameters are found to be significantly different: *Sq* (p = 0.002), *Ssk* (p = 0.021), *Sku* (p = 0.020), *Sp* (p = 0.050), *Sv* (p = 0.017), *Sz* (p = 0.023), *Sa* (p = 0.002),  *Smc* (p = 0.003), *Sxp* (p = 0.001)**,** *Spc* (p = 0.017), *S10z, S5p, S5v* (p = 0.038), Sda, Sha, Sdv, Shv, Sk (p = 0.007), *Spk* (p = 0.003), *Svk* (p = 0.003), *Height mean* (p = 0.027), *SRC* (p =

0.028), and *Reg coeff* (p = 0.011). It is notable that of the few parameters that are not significantly different between histpathologically diagnosed keratoses and those not diagnosed as being keratoses, most are volumetric parameters (like *Vm, Vv, Vmp, Vmc, Vvc, Vvc,* and *Vvv*).

## 4 | DISCUSSION

Haptics designed for simulating skin texture might benefit from these more subtle characterizations of skin lesion texture that are based more on the mathematical shapes of disease, rather than familiarity that is inherently subjectively defined by palpation and experience.[43] These strides are already taking place with laser scanned skin and simple measures of roughness.[44] For instance, one study used modeling tools to investigate different features of texture of seborrheic keratosis and melanoma on the basis of two-dimensional images, creating a classification system that could correctly differentiate seborrheic keratosis lesion images from other pigmented lesions.[49] Another system relied on integrated digital dermoscopy analysis to gain objective measurements to contribute to computer-aided melanoma diagnosis, evaluating geometries, colors, textures, and islands of colors across variables.[50] Similarly, through further investigation of smaller details and more complex measures, such as those outlined in our study, we may find ways to push beyond digital representations of palpation and achieve systems in which surface metrology of skin is measured and analyzed in an automated manner akin to the way dermoscopic images are being used for cutaneous malignancies, including melanoma.[45] The hopes for automated detection of skin pathologies may soon be realized.[46]

Our results suggest that texture features, as measured through a variety of roughness parameters, can effectively classify seborrheic keratoses. Even though there was no statistically sound difference measured between SK center versus SK border, it is possible that this is due to the heterogeneity of the surface seborrheic keratoses. Not one seborrheic keratosis is identical to another, despite sharing a multitude of diagnostic features on visual inspection with the naked eye, dermoscopy, and other modalities. Even at the histologic levels, there are several subtypes of seborrheic keratoses that can be identified: hyperkeratotic type, acanthotic type, reticular/adenoid type, clonal type, irritated type, and more.[51] In our study, it is possible that there was variability in the region of surface scanned. For example, one scan for one SK lesion's center could have been taken from a more rounded, crusted surface while another scan for another SK lesion's center could have been flatter and less warty. Similarly, no statistically significant difference in SK parameters on the basis of location could have also been influenced by the range of SK clinical heterogeneity overall.

### 4.1 | Correlating keratoses with surface metrology parameters with level of sun exposure

Several statistically significant differences in measured roughness parameters were identified in seborrheic keratoses as a result of level of sun exposure (categorized as none versus low level versus high level) and sex (male versus female). Age is a known risk factor for the development of SKs, exhibiting a higher prevalence in older populations as well as a tendency to increase in number with age. Recent studies of dermatology patients in Brazil and Australia found strong evidence of age-related prevalence, demonstrating higher average numbers of SKs in the elderly.[53-55] Aging skin is particularly due to chronic ultraviolet (UV) exposure. Potential underlying contributors include oncogenic mutations, but the mechanism through which this occurs is still unclear. Chronic sun exposure has been implicated in the pathogenesis of cutaneous malignancies, including squamous cell carcinoma and basal cell carcinoma.[56] Some suggest that prolonged UV exposure may also contribute to SK formation, but the literature has not clarified the causal role in SK development. For instance, a Korean study demonstrated how both age and cumulative sun exposure are independent contributory factors involved in the development of SKs.[58] A separate Dutch study revealed that painful sunburns (a direct result of excess sun exposure) resulted in an increased risk of SK.[57] Similarly, our results indicated significant differences in roughness parameters of seborrheic keratoses as related to sun exposure, suggesting that there may be some texture variability as related to UV exposure, as opposed to location which did not produce statistically significant differences in our study.

4.2 | Correlating keratoses with surface metrology parameters with histopathology

Topographical characteristics are of key significance in the differential diagnosis of SKs from other cutaneous lesions or neoplasms. Histopathological examination is still relevant in the diagnosis of SK, particularly in the cases of atypical presentation when clinical appearance is not conclusive.[35] It identifies specific features such as hyperkeratosis, basaloid cell proliferation, and pseudo-horn or milia-like cysts, while microscopic examination frequently reveals epidermal alterations and pseudocysts.[26, 35-36] For instance, a notable case in the literature emphasized the critical role of histopathology in confirming the diagnosis of SK in a lesion that mimicked squamous cell carcinoma.[35]

The results obtained in our study show that histologic diagnosis is the most definitive feature that distinguishes seborrheic keratosis lesions from normal skin or other non-keratotic lesions. This could be concluded due to the presence of greater than twenty measured roughness parameters that showed statistically significant differences between keratotic and normal skin. Histopathology, which is arguably the standard for dermatologic diagnosis in many scenarios, requires a biopsy. Our study is the first to date to link surface texture, which does not require a biopsy for surface metrology assessment, with histopathology, showing promising avenues as a complementary noninvasive diagnostic modality.

4.3 | Clinical relevance & Further Steps

The introduction of noninvasive diagnostic modalities has enhanced the identification and treatment of seborrheic keratoses (SKs) to a considerable extent. However, findings from recent studies highlight the need for reliable quantitative data in the diagnosis of SKs.[28,33] For instance, the analysis of a giant perigenital SK underscored the importance of accurate diagnosis and preventing mismanagement.[36]

The clinical implications of our findings are significant, illustrating how advanced imaging techniques allow for accurate, standardized diagnosis of SKs without the need for more invasive methods. Not only does this diminish the need to perform a biopsy, but it also minimizes patient-centered fear surrounding invasive procedures to begin with, positively impacting patient outcomes.[37] Surface metrology allows one to quantify SKs by an entirely objective method, eliminating the need for specialized training in visual interpretation. After all, visual inspection of pigmented lesions, including SKs, is prone to inter-observer and intra-observer variability.[60] Thus, a more advanced imaging diagnostic could assist dermatologists and primary care physicians in making a uniform diagnosis, without entirely replacing their role in diagnosis by visual inspection.[42] It could enhance the automatic classification schemes that exist through a less subjective, more quantitative approach. After all, there are consequences of misclassifying melanoma as benign is much greater than mistakenly identifying a benign tumor like a seborrheic keratosis as malignant. This poses the risk of delayed diagnosis and improper disease management, affecting mortality rates and patient outcomes, especially in melanoma management.[61-63]

If surface metrology is applied in the future to other pigmented lesions, this may offer a step towards curbing the mortality caused by particular cutaneous lesions, especially in scenarios where pigmented lesions appear clinically and dermoscopically equivocal to malignancy.[34] However, we need to build a better understanding of this on greater population studies on measuring surface roughness in other histologically diagnosed skin lesions warranting differentiation. If we can use surface metrology to distinguish healthy skin from SKs, perhaps this is an opportunity to investigate how to differentiate healthy skin and skin malignancies through such quantification. And, if these surface texture parameters consistently differ between groups, this offers an avenue for diagnosis without performing an invasive biopsy. This advancement in our study underscores the importance of continued research and innovation in dermatologic diagnostic modalities.

5 | CONCLUSIONS

This study demonstrates that there are quantifiable and predictable patterns of surface textures that can be compared between keratotic and normal skin surfaces. These findings are most easily interpreted by observing significant differences in roughness parameters between keratotic and normal skin, whether that be normal control skin or non-keratotic, in terms of sex, degree of sun exposure, and histologic diagnosis. This is the first study to investigate and effectively demonstrate the quantification of texture in dermatology by quantifying surface texture parameters of a very clinically prevalent benign lesion, seborrheic keratoses. The parameters investigated with the optical profiler may be used in future investigations to further evaluate the diagnostic precision of this method on a broader range of benign or malignant pigmented lesions, especially basal cell carcinoma and melanomas, both of which keratoses can clinically mimic upon visual inspection. Nonetheless, these findings suggest that this method has the potential to be used as a noninvasive complementary adjunct to current noninvasive diagnostic modalities in dermatology.

| TABLE

TABLE 1. Parameters quantified by the optical white light profiler

| Name | Unit | Context | Description |
|---|---|---|---|
| **ISO 25718-2** | | | |
| **Height parameters** | | | |
| Sq | µm | | Root-mean-square height |
| Ssk | <no unit> | | Skewness |
| Sku | <no unit> | | Kurtosis |
| Sp | µm | | Maximum peak height |
| Sv | µm | | Maximum pit height |
| Sz | µm | | Maximum height |
| Sa | µm | | Arithmetic mean height |
| **Hybrid parameters** | | | |
| Smr | % | c = 1 µm under the highest peak | Areal material ratio |
| Smc | µm | p = 10% | Inverse areal material ratio |
| Sxp | µm | p = 50%  q = 97.5% | Extreme peak height |
| **Functional parameters** | | | |
| Sal | µm | s = 0.2 | Autocorrelation length |
| Str | <no unit> | s = 0.2 | Texture-aspect ratio |
| Std | ° | Reference angle = 0° | Texture direction |
| **Feature parameters** | | | |
| Sdq | <no unit> | | Root-mean-square gradient |
| Sdr | % | | Developed interfacial area ratio |
| **Functional parameters (Volume)** | | | |
| Vm | µm³/µm² | p = 10% | Material volume |
| Vv | µm³/µm² | p = 10% | Void volume |
| Vmp | µm³/µm² | p = 10% | Peak material volume |
| Vmc | µm³/µm² | p = 10%  q = 80% | Core material volume |
| Vvc | µm³/µm² | p = 10%  q = 80% | Core void volume |
| Vvv | µm³/µm² | p = 80% | Pit void volume |
| **Feature parameters** | | | |
| Spd | 1/µm² | pruning = 5% | Density of peaks |
| Spc | 1/µm | pruning = 5% | Arithmetic mean peak curvature |
| S10z | µm | pruning = 5% | Ten point height |
| S5p | µm | pruning = 5% | Five point peak height |
| S5v | µm | pruning = 5% | Five point pit height |
| Sda | µm² | pruning = 5% | Mean dale area |
| Sha | µm² | pruning = 5% | Mean hill area |
| Sdv | µm³ | pruning = 5% | Mean dale volume |
| Shv | µm³ | pruning = 5% | Mean hill volume |
| **Functional parameters (Stratified surfaces)** | | | |
| Sk | µm | Gaussian filter  0.08 mm | Core roughness depth |
| Spk | µm | Gaussian filter  0.08 mm | Reduced summit height |
| Svk | µm | Gaussian filter  0.08 mm | Reduced valley depth |
| Smr1 | % | Gaussian filter  0.08 mm | Upper bearing area |
| Smr2 | % | Gaussian filter  0.08 mm | Lower bearing area |
| Spq | <no unit> | Gaussian filter  0.08 mm | Plateau root-mean-square roughness |
| Svq | <no unit> | Gaussian filter  0.08 mm | Valley root-mean-square roughness |
| Smq | <no unit> | Gaussian filter  0.08 mm | Material ratio at plateau-to-valley transition |
| **Texture direction** | | | |
| Isotropy | % | | Uniformity of orientation of tiles |
| First Direction | ° | | Most common orientation of tiles |
| Second Direction | ° | | Second-most common orientation of tiles |
| Third Direction | ° | | Third-most common orientation of tiles |
| **Motifs analysis** | | | |
| Number of motifs | <no unit> | | Number of recognized local peaks delimited by slope inflections |
| Height[Mean] | µm | | Mean height of motifs |
| Area[Mean] | µm² | | Mean area of motifs |
| **Scale-Sensitive Fractal Analysis (SSFA)** | | | |
| Y Max | <no unit> | | The size of tiles at the finest scale |
| SRC threshold | <no unit> | | Threshold relative length of the smooth-rough crossover |
| Smooth-rough crossover (SRC) | µm | | Scale of the smooth-rough crossover |
| Reg. coefficient R² | <no unit> | | A measure of the accuracy of the complexity value |
| Fractal complexity (Lsfc) | <no unit> | | How much of the surface is more complex than a Euclidean plane |
| Fractal dimension (Dls) | <no unit> | | The range of scales in which the slope of the line is straight |
| Scale of max complexity (Smfc) | µm | | Scale where highest complexity is found |

Table 1 demonstrates parameter variables and their respective abbreviations and short descriptions that are quantified by the optical white light profiler and included in the study's statistical analysis.

| FIGURES

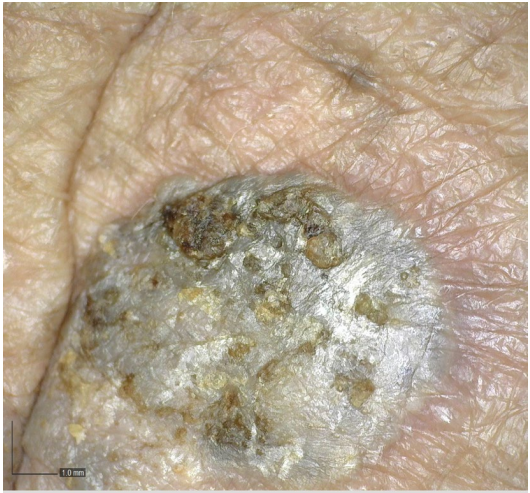

**FIGURE 1 (\*\*INCLUDE .TIFF\*\*)**
This figure represents the image of a suspected seborrheic keratosis lesion from the sample.

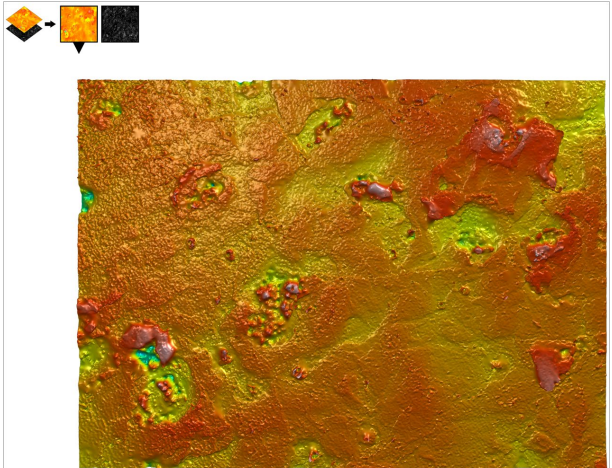

**FIGURE 2 (\*\*INCLUDE .TIFF\*\*)**
Specimen casts were scanned with Sensofar S neox optical profiler. This figure illustrates the variations in surface texture seen on surface metrology.

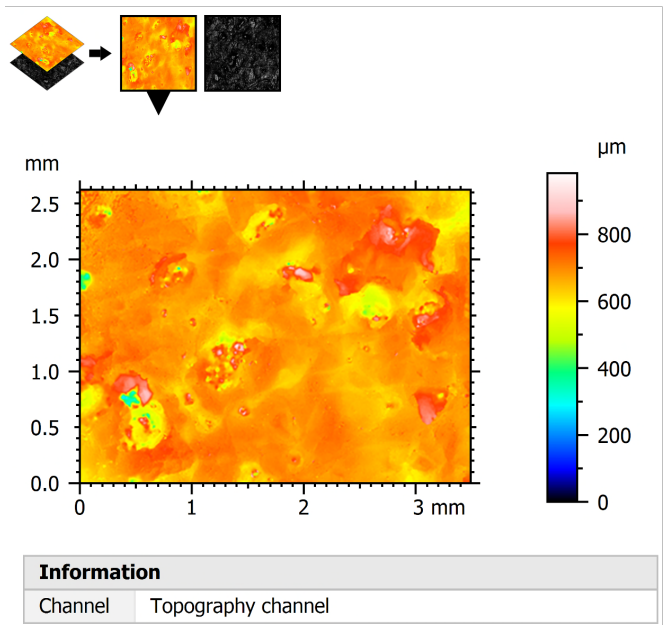

**FIGURE 3 (**INCLUDE .TIFF**)**

This figure summarizes the final scan from which subsequent data was extracted (ISO ISO-25178-2 parameters, texture direction, motifs analysis, and scale-sensitize fractal analysis).

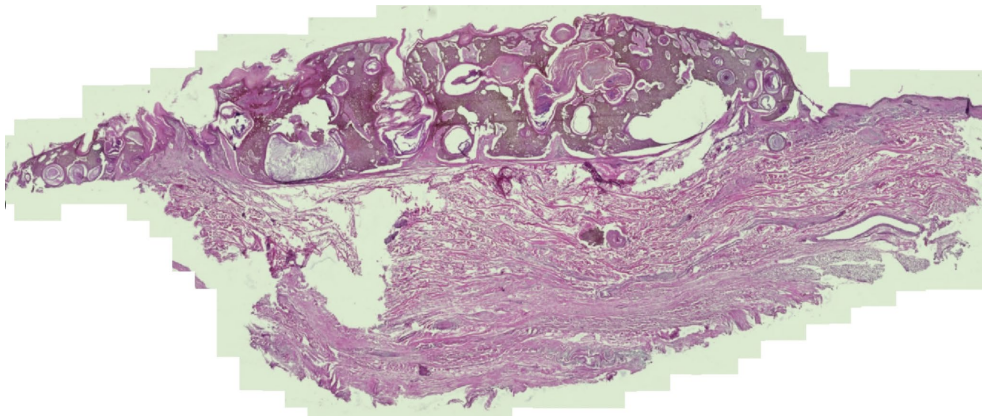

**FIGURE 4 (**INCLUDE .TIFF**)**

This figure is an example of a seborrheic keratosis lesion that was confirmed by review by the dermatopathologist based on key histological features.

## | APPENDIX

S1. The complete dataset of parameters derived from scans. This supporting file will be submitted separately.